\documentclass[twocolumn,showpacs,preprintnumbers,amsmath,aps]{revtex4}
\usepackage{graphicx}
\usepackage{dcolumn}
\usepackage{bm}
\usepackage{color}
\begin{document}
\newcommand{\kvec}{\mbox{{\scriptsize {\bf k}}}}
\def\eq#1{(\ref{#1})}
\def\fig#1{Fig.\hspace{1mm}\ref{#1}}
\def\tab#1{\hspace{1mm}\ref{#1}}
\title{
---------------------------------------------------------------------------------------------------------------\\
On the critical temperature and the energy gap in dense SiH$_4$(H$_2$)$_2$ at 250 GPa}
\author{R. Szcz{\c{e}}{\'s}niak, A.P. Durajski}
\affiliation{Institute of Physics, Cz{\c{e}}stochowa University of Technology, Al. Armii Krajowej 19, 42-200 Cz{\c{e}}stochowa, Poland}
\email{adurajski@wip.pcz.pl}
\date{\today} 
\begin{abstract}
The critical temperature ($T_{C}$) and the energy gap ($2\Delta\left(T\right)$) for the superconductor SiH$_4$(H$_2$)$_2$ at $250$ GPa 
have been calculated. The wide range of the Coulomb pseudopotential's values has been considered: $\mu^{\star}\in\left<0.1,0.3\right>$. It has been stated that $T_{C}$ decreases together with the increase of $\mu^{\star}$ from $129.83$ K to $81.40$ K. The low-temperature energy gap ($T\sim 0$ K) decreases together with the increase of the Coulomb pseudopotential from $50.96$ meV to $30.12$ meV. The high values of $2\Delta\left(0\right)$ mean that the dimensionless ratio $R_{\Delta}\equiv 2\Delta\left(0\right)/k_{B}T_{C}$ significantly exceeds the value predicted by the classical BCS theory. In the considered case: $R_{\Delta}\in\left<4.55,4.29\right>$. Due to the unusual dependence of the critical temperature and the energy gap on $\mu^{\star}$, the analytical expressions for $T_{C}\left(\mu^{\star}\right)$ and $\Delta\left(\mu^{\star}\right)$ have been given.
\end{abstract}
\maketitle

\vspace*{1cm}


The high-pressure superconducting state, induced by the electron-phonon interaction, may have the very high value of the critical temperature. Due to the small mass of the nuclei forming the crystal lattice and the lack of the external electron shells, the most interesting properties should have been revealed by the high-pressure superconducting phase in the hydrogen \cite{Ashcroft}, \cite{Maksimov}, \cite{Szczesniak1dodatkowa}, \cite{McMahon}. 

The {\it ab initio} calculations suggest that the metallization of the hydrogen takes place for the pressure $p\sim 400$ GPa \cite{Stadele}. In the considered case, the hydrogen exists in the molecular phase. It has been stated that the critical temperature in the molecular hydrogen reaches high values ($\sim 240$ K for $p=450$ GPa); the other thermodynamic parameters are significantly different from the predictions of the classical BCS theory \cite{Cudazzo}, \cite{Szczesniak1}, \cite{Szczesniak2}. Above the pressure equal to $500$ GPa follows the dissociation of the hydrogen's molecular phase into the atomic phase \cite{Stadele}, \cite{Natoli}, \cite{Johnson}, \cite{Pickard}. The superconducting state in the atomic phase has been analyzed up to the value of $3.5$ TPa \cite{McMahon}. The obtained results suggest the extremely high critical temperature ($\sim 600$ K for $p=2$ TPa) \cite{Maksimov}. In the considered case, the ratio of the low-temperature energy gap and the critical temperature equals more than $6$, so it is comparable with the values measured in the cuprates \cite{Szczesniak1dodatkowa}, \cite{Szczesniak3}.

Because of the high metallization's pressure, the experimental analysis of the hydrogen's superconducting state is unavailable. For this reason, currently examined is the existence of the high-temperature superconducting state in the compounds of IV group: ${\rm CH_{4}}$, ${\rm GeH_{4}}$, ${\rm SnH_{4}}$, and ${\rm SiH_{4}}$ \cite{Ashcroft1}, \cite{Tse}, \cite{Gao}, \cite{Canales}, \cite{Gao1}, \cite{Chen}, \cite{Eremets}. 

In the case of ${\rm SiH_{4}}$, the metallization occurs at the pressure $50-60$ GPa \cite{Chen}, \cite{Eremets}. The low-temperature superconducting state ($T_{C}\simeq 17$ K) has been observed for the pressure at $96$ GPa and $120$ GPa \cite{Eremets}.  It is possible that the higher values of the critical temperature would be measured in the range of the higher pressures or in the compounds of the silicon like: ${\rm Si_{2}H_{6}}$ or ${\rm SiH_{4}\left(H_{2}\right)_{2}}$ \cite{Jin}, \cite{Li}.

In the case of ${\rm SiH_{4}\left(H_{2}\right)_{2}}$, it has been stated that the critical temperature can take the value of $107$ K for the pressure at 250 GPa \cite{Li}. We note that $T_{C}$ has been estimated on the basis of the McMillan formula \cite{McMillan}.

From the microscopic point of view, the growth of the critical temperature in ${\rm SiH_{4}\left(H_{2}\right)_{2}}$ is related to the strong interaction between the additional molecules ${\rm H_{2}}$ and ${\rm SiH_{4}}$, which leads to the significant increase of the electron-phonon coupling constant. 

In the present study, we have determined the exact dependence of the critical temperature and the energy gap on the Coulomb pseudopotential for ${\rm SiH_{4}\left(H_{2}\right)_{2}}$ compound ($p=250$ GPa). Due to the very high value of the electron-phonon coupling constant ($\lambda=1.61$), the calculations have been carried out by strictly solving the Eliashberg equations \cite{Eliashberg}.


The Eliashberg equations on the imaginary axis assume the following form:
\begin{equation}
\label{r1}
\phi_{n}=\frac{\pi}{\beta}\sum_{m=-M}^{M}
\frac{\lambda\left(i\omega_{n}-i\omega_{m}\right)-\mu^{\star}\theta\left(\omega_{c}-|\omega_{m}|\right)}
{\sqrt{\omega_m^2Z^{2}_{m}+\phi^{2}_{m}}}\phi_{m},
\end{equation}
\begin{equation}
\label{r2}
Z_{n}=1+\frac{1}{\omega_{n}}\frac{\pi}{\beta}\sum_{m=-M}^{M}
\frac{\lambda\left(i\omega_{n}-i\omega_{m}\right)}{\sqrt{\omega_m^2Z^{2}_{m}+\phi^{2}_{m}}}\omega_{m}Z_{m}.
\end{equation}

The quantity $\phi_{n}\equiv\phi\left(i\omega_{n}\right)$ represents the order parameter function; $Z_{n}\equiv Z\left(i\omega_{n}\right)$ denotes the wave function renormalization factor; $n$-th Matsubara frequency is defined by the expression: 
$\omega_{n}\equiv \left(\pi / \beta\right)\left(2n-1\right)$, where $\beta\equiv\left(k_{B}T\right)^{-1}$, and $k_{B}$ is the Boltzmann constant. In the Eliashberg formalism, the order parameter is given by the formula: $\Delta\equiv \phi/Z$.

The symbol $\lambda\left(z\right)$ denotes the pairing kernel for the electron-phonon interaction:
\begin{equation}
\label{r3}
\lambda\left(z\right)\equiv 2\int_0^{\Omega_{\rm{max}}}d\Omega\frac{\Omega}{\Omega ^2-z^{2}}\alpha^{2}F\left(\Omega\right),
\end{equation}
where $\alpha^{2}F\left(\Omega\right)$ is the Eliashberg function. For ${\rm SiH_{4}\left(H_{2}\right)_{2}}$ under the pressure at 250 GPa, the Eliashberg function has been determined in the paper \cite{Li}. The maximum phonon frequency is equal to $375.98$ meV.

The depairing Coulomb interaction is parameterized with the help of the Coulomb pseudopotential $\mu^{\star}$. The symbol $\theta$ denotes the Heaviside function and $\omega_{c}$ is the cut-off frequency; $\omega_{c}=3\Omega_{\rm{max}}$.

The Eliashberg equations have been solved for $2201$ Matsubara frequencies ($M = 1100$). The numerical methods used in the paper have been presented in: \cite{Szczesniak4}, \cite{Szczesniak5}, \cite{Szczesniak6}, and \cite{Szczesniak7}. The convergence of the solutions has been obtained for $T\geq T_{0}=11.6$ K.

Using the Eliashberg equations on the imaginary axis one can precisely calculate the value of the critical temperature in the dependence on the Coulomb pseudpotential. In the study, it has been assumed $\mu^{\star}\in\left<0.1,0.3\right>$; due to the fact that the exact value of the Coulomb pseudopotential is very difficult to calculate with the use of the {\it ab initio} methods.

In order to estimate the physical value of the order parameter, the Eliashberg equations should be solved in the mixed representation (the equations are defined simultaneously on the imaginary and real axis) \cite{Marsiglio}:

%
\begin{widetext}
\begin{eqnarray}
\label{r4}
\phi\left(\omega+i\delta\right)&=&
                                  \frac{\pi}{\beta}\sum_{m=-M}^{M}
                                  \left[\lambda\left(\omega-i\omega_{m}\right)-\mu^{\star}\theta\left(\omega_{c}-|\omega_{m}|\right)\right]
                                  \frac{\phi_{m}}
                                  {\sqrt{\omega_m^2Z^{2}_{m}+\phi^{2}_{m}}}\\ \nonumber
                              &+& i\pi\int_{0}^{+\infty}d\omega^{'}\alpha^{2}F\left(\omega^{'}\right)
                                  \left[\left[N\left(\omega^{'}\right)+f\left(\omega^{'}-\omega\right)\right]
                                  \frac{\phi\left(\omega-\omega^{'}+i\delta\right)}
                                  {\sqrt{\left(\omega-\omega^{'}\right)^{2}Z^{2}\left(\omega-\omega^{'}+i\delta\right)
                                  -\phi^{2}\left(\omega-\omega^{'}+i\delta\right)}}\right]\\ \nonumber
                              &+& i\pi\int_{0}^{+\infty}d\omega^{'}\alpha^{2}F\left(\omega^{'}\right)
                                  \left[\left[N\left(\omega^{'}\right)+f\left(\omega^{'}+\omega\right)\right]
                                  \frac{\phi\left(\omega+\omega^{'}+i\delta\right)}
                                  {\sqrt{\left(\omega+\omega^{'}\right)^{2}Z^{2}\left(\omega+\omega^{'}+i\delta\right)
                                  -\phi^{2}\left(\omega+\omega^{'}+i\delta\right)}}\right],
\end{eqnarray}
and
\begin{eqnarray}
\label{r5}
Z\left(\omega+i\delta\right)&=&
                                  1+\frac{i}{\omega}\frac{\pi}{\beta}\sum_{m=-M}^{M}
                                  \lambda\left(\omega-i\omega_{m}\right)
                                  \frac{\omega_{m}Z_{m}}
                                  {\sqrt{\omega_m^2Z^{2}_{m}+\phi^{2}_{m}}}\\ \nonumber
                              &+&\frac{i\pi}{\omega}\int_{0}^{+\infty}d\omega^{'}\alpha^{2}F\left(\omega^{'}\right)
                                  \left[\left[N\left(\omega^{'}\right)+f\left(\omega^{'}-\omega\right)\right]
                                  \frac{\left(\omega-\omega^{'}\right)Z\left(\omega-\omega^{'}+i\delta\right)}
                                  {\sqrt{\left(\omega-\omega^{'}\right)^{2}Z^{2}\left(\omega-\omega^{'}+i\delta\right)
                                  -\phi^{2}\left(\omega-\omega^{'}+i\delta\right)}}\right]\\ \nonumber
                              &+&\frac{i\pi}{\omega}\int_{0}^{+\infty}d\omega^{'}\alpha^{2}F\left(\omega^{'}\right)
                                  \left[\left[N\left(\omega^{'}\right)+f\left(\omega^{'}+\omega\right)\right]
                                  \frac{\left(\omega+\omega^{'}\right)Z\left(\omega+\omega^{'}+i\delta\right)}
                                  {\sqrt{\left(\omega+\omega^{'}\right)^{2}Z^{2}\left(\omega+\omega^{'}+i\delta\right)
                                  -\phi^{2}\left(\omega+\omega^{'}+i\delta\right)}}\right]. 
\end{eqnarray}
\end{widetext}
%

The symbol $N\left(\omega\right)$ and $f\left(\omega\right)$ denotes the Bose-Einstein function and the Fermi-Dirac function, respectively. The equations \eq{r4} and \eq{r5} have been solved with the help of the numerical methods used in the papers: \cite{Szczesniak8}, \cite{Szczesniak9} and \cite{Szczesniak10}.

\begin{figure*}[th]
\includegraphics[scale=0.60]{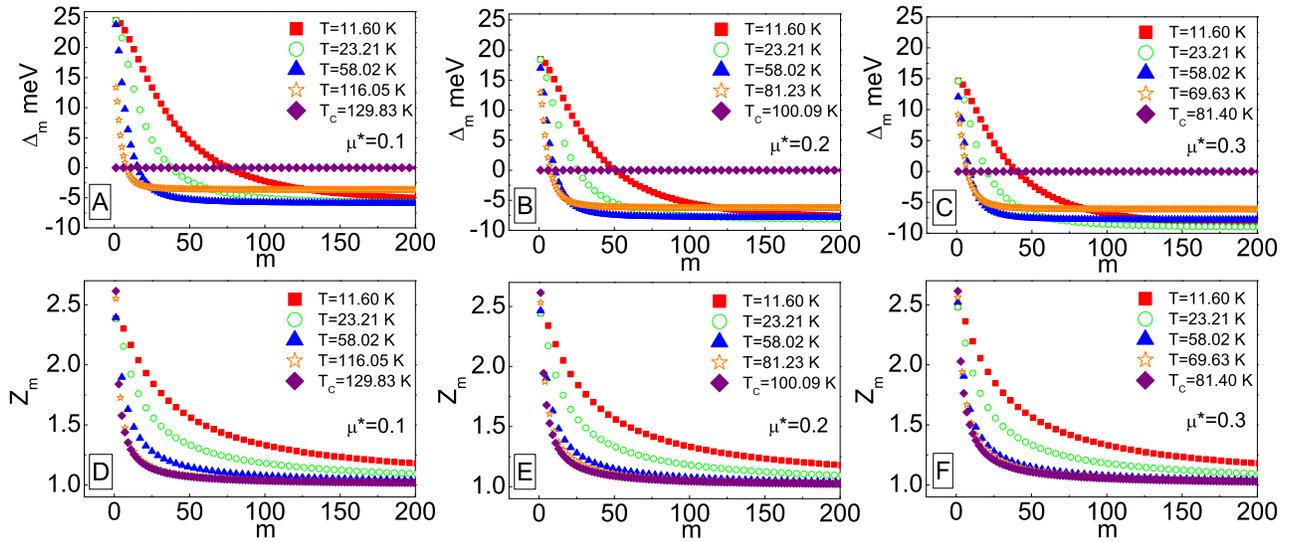}
\caption{
The dependence of the order parameter (figures (A)-(C)) and the wave function renormalization factor (figures (D)-(F)) on the number $m$ for the selected values of the temperature and the Coulomb pseudopotential. The first $200$ values of the functions $\Delta_{m}$ and $Z_{m}$ have been presented.}
\label{f1}
\end{figure*}

The solutions of the Eliashberg equations on the imaginary axis have been presented in \fig{f1}. In particular, in \fig{f1} (A)-(C) there is plotted the dependence of the order parameter on the number $m$ for the selected values of the temperature and the Coulomb pseudopotential. On the other hand, in \fig{f1} (D)-(F) there is presented the influence of the temperature and the Coulomb pseudopotential on the form of the wave function renormalization factor.  

%
%

On the basis of the obtained data, it has been found that the increase of the temperature causes the significant decrease of the maximum value of the order parameter ($\Delta_{m=1}$); also decreases the half-width of the considered function. In the case of the increase of the Coulomb pseudopotential, the function of the order parameter strongly decreases in the range of the lower Matsubara frequencies and becomes saturated at the level of the progressively smaller values.

Analyzing the results obtained for the wave function renormalization factor, the very weak dependence of the function $Z_{m}$ on $T$ and $\mu^{\star}$ can be easily observed. From the physical point of view, the achieved result indicates that the electron effective mass is slightly responsive to the changes in the values of the temperature or $\mu^{\star}$ parameter.

%
\begin{figure}[th]
\includegraphics[width=\columnwidth]{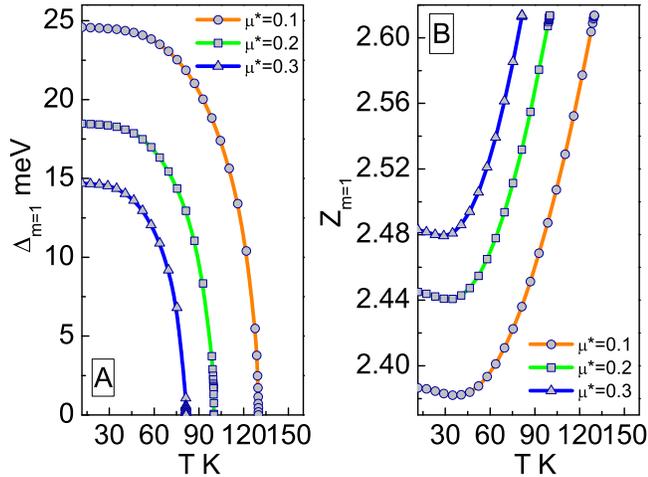}
\caption{
The maximum value of the order parameter (figure A) and the maximum value of the wave function renormalization factor (figure B) as the function of the temperature for the selected values of the Coulomb pseudopotential.}
\label{f2}
\end{figure}
%

The temperature dependence of the order parameter and the wave function renormalization factor can be traced in the most convenient way by plotting the courses of the functions $\Delta_{m=1}\left(T\right)$ and $Z_{m=1}\left(T\right)$. The results have been presented in  \fig{f2} (A) and (B). In the case of the order parameter, the quantity $2\Delta_{m=1}\left(T\right)$ with the good approximation reproduces the temperature evolution of the energy gap at the Fermi level. Let us notice that the shape of the function $\Delta_{m=1}\left(T\right)$ can be parameterized with the help of the following expression: $\Delta_{m=1}\left(T,\mu^{\star}\right)=\Delta_{m=1}\left(T_{0},\mu^{\star}\right)\sqrt{1-\left(\frac{T}{T_{C}}\right)^{\beta}}$, where $\Delta_{m=1}\left(T_{0},\mu^{\star}\right)=116\left(\mu^{\star}\right)^{2}-95.9\mu^{\star}+33.01$ meV and $\beta=3.4$.

In the case of the second solution of the Eliashberg equations, \fig{f2} (B) confirms the very weak dependence of $Z_{m=1}$ on the temperature and the Coulomb pseudopotential. Note that for $T=T_{C}$, the maximum value of the wave function renormalization factor is independent of $\mu^{\star}$. In the considered case, it can be calculated with the help of the simple formula: $\left[Z_{m=1}\right]_{T=T_{C}}=1+\lambda=2.61$. The identical value has been achieved on the basis of the numerical calculations. The above result proves the high accuracy of the used numerical methods.

\begin{figure}%
\includegraphics[width=\columnwidth]{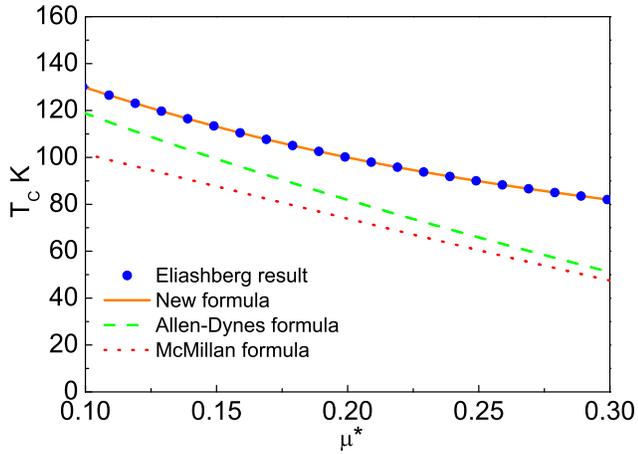}
\caption{
The critical temperature as the function of the Coulomb pseudopotential.}
\label{f3}
\end{figure}

Below, there is determined the strict dependence of the critical temperature on the Coulomb pseudopotential. In the considered case, the shape of the function $T_{C}\left(\mu^{\star}\right)$ has been reproduced numerically for $300$ values of $\mu^{\star}$ from the range $0.1$ - $0.3$; the condition $\left[\Delta_{m=1}\right]_{T=T_{C}}=0$ has been used. The results have been presented in \fig{f3}. In addition, there are also plotted the critical temperature obtained using the classical McMillan formula and the Allen-Dynes expression \cite{McMillan}, \cite{AllenDynes}. It is easy to see that the critical temperature calculated in the analytical way is underestimated, especially in the range of the higher values of $\mu^{\star}$. However, the Allen-Dynes expression much better predicts $T_{C}$ than McMillan formula. 

In view of the difficulties in precise estimating the value of the critical temperature by the classical expressions, the new formula for $T_{C}$ has been given. In particular, we have used the method of least squares and $300$ exact values of the function $T_{C}\left(\mu^{\star}\right)$. The obtained result takes the form: 
\begin{equation}
\label{r6}
k_{B}T_{C}=f_{1}f_{2}\frac{\omega_{{\rm ln}}}{1.37}\exp\left[\frac{-1.125\left(1+\lambda\right)}{\lambda-\mu^{\star}}\right],
\end{equation}
where $f_{1}$ and $f_{2}$ denote the correction functions \cite{AllenDynes}: 
$f_{1}\equiv\left[1+\left(\frac{\lambda}{\Lambda_{1}}\right)^{\frac{3}{2}}\right]^{\frac{1}{3}}$ and
$f_{2}\equiv 1+\frac{\left(\frac{\sqrt{\omega_{2}}}{\omega_{\rm{ln}}}-1\right)\lambda^{2}}{\lambda^{2}+\Lambda^{2}_{2}}$. The quantities 
$\Lambda_{1}$ and $\Lambda_{2}$ have the form:
$\Lambda_{1}=2-0.14\mu^{\star}$ and $\Lambda_{2}=\left(0.27+10\mu^{\star}\right)\left(\sqrt{\omega_{2}}/\omega_{\ln}\right)$. 
The parameters $\omega_{{\rm ln}}$ and $\sqrt{\omega_{2}}$ are equal to $71.76$ meV and $122.67$ meV, respectively. 

On the basis of \fig{f3}, it is easy to see that the formula \eq{r6} exactly reproduces the numerical values of the critical temperature.

%
%


The solutions of the Eliashberg equations on the imaginary axis are used as the input parameters to the Eliashberg equations in the mixed representation. Those equations allow to continue analytically the order parameter and the wave function renormalization on the real axis: $\Delta_{m}\rightarrow \Delta\left(\omega\right)$ and $Z_{m}\rightarrow Z\left(\omega\right)$.

\begin{figure*}[th]
\includegraphics[scale=0.55]{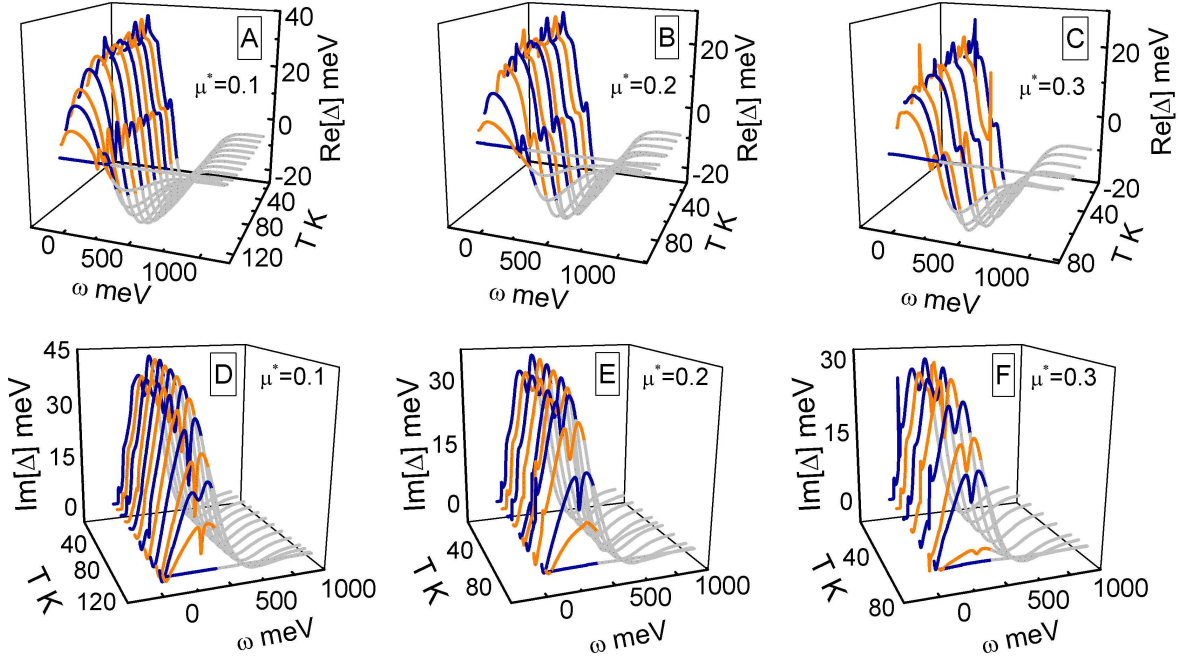}
\caption{
The real part of the order parameter (figures (A)-(C)) and the imaginary part of the order parameter (figures (D)-(F)) on the real axis for the selected values of the temperature and the Coulomb pseudopotential. The blue and orange lines represent the solutions for $\omega\in\left<0,\Omega_{\rm max}\right>$, whereas the grey lines correspond to the solutions for $\omega\in\left(\Omega_{\rm max}, \omega_{c}\right>$.} 
\label{f4}
\end{figure*}

In \fig{f4}, the results obtained for the order parameter have been presented. In particular, in \fig{f4} (A)-(C) we have plotted the courses of the real part of the function $\Delta\left(\omega\right)$ for the selected values of the temperature and the Coulomb pseudopotential; $\omega\in\left<0,\omega_{c}\right>$. It has been stated that in the range of the frequencies from $0$ to $\Omega_{{\rm max}}$ the function 
${\rm Re}\left[\Delta\left(\omega\right)\right]$ has the complicated shape, which can be divided into the number of the local maxima and minima (especially in the range of the low temperatures). It should be emphasized that for $\omega\in\left<0,\Omega_{{\rm max}}\right>$ the complex behavior of the real part of the order parameter is induced by the complicated shape of the Eliashberg function. Analyzing \fig{f4} (A)-(C), it can be additionally noticed that the increase of the temperature smooths the course of the function ${\rm Re}\left[\Delta\left(\omega\right)\right]$. In the range of the higher frequencies ($\omega\in\left(\Omega_{{\rm max}},\omega_{c}\right>$), the real part of the order parameter is subjected to the saturation.

The imaginary part of the order parameter on the real axis is plotted in \fig{f4} (D)-(F). There are clear similarities between the behavior of the function ${\rm Im}\left[\Delta\left(\omega\right)\right]$ and ${\rm Re}\left[\Delta\left(\omega\right)\right]$. In particular, in the range of the frequencies from $0$ to $\Omega_{{\rm max}}$, the function ${\rm Im}\left[\Delta\left(\omega\right)\right]$ is characterized by the complicated course, closely correlated with the shape of the Eliashberg function. For the higher frequencies it also becomes saturated.

\begin{figure*}[th]
\includegraphics[scale=0.60]{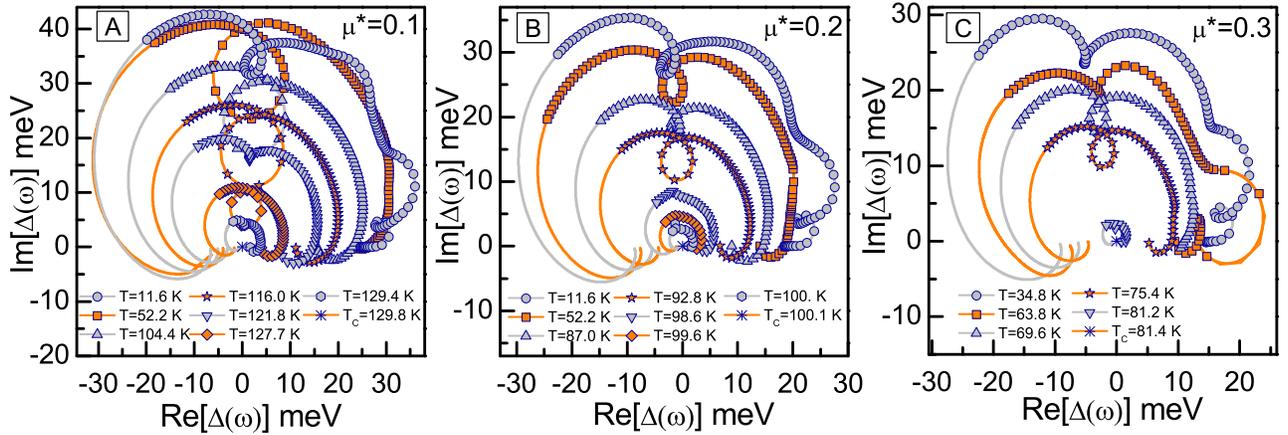}
\caption{
(A)-(C) The order parameter on the complex plane for the selected values of the temperature and the Coulomb pseudopotential. The lines with symbols represent the solutions for $\omega\in\left<0,\Omega_{\rm max}\right>$, whereas the regular lines correspond to the solutions for $\omega\in\left(\Omega_{\rm max}, \omega_{c}\right>$.} 
\label{f5}
\end{figure*}

In order to perform the thorough analysis, the function $\Delta\left(\omega\right)$ should be plotted on the complex plane (\fig{f5} (A)-(C)). After making the corresponding transformations, the characteristic spiral forms have been obtained. On the basis of the presented results, the complicated structure of the order parameter for the frequencies from $0$ to $\Omega_{{\rm max}}$ can be seen very clearly. 

The obtained results allow to identify the values of the frequency for which the effective potential of the electron-electron interaction is attractive 
(${\rm Re}\left[\Delta\left(\omega\right)\right]>0$) \cite{Varelogiannis}. In the considered case these may be two compartments: the first one extends from $0$ to $\omega_{p}<\Omega_{\rm max}$ and exists up to the critical temperature and for $\mu^{\star}\in\left<0.1,0.3\right>$. It should be strongly emphasized that the increase of the temperature and the Coulomb pseudopotential causes the clear decrease of $\omega_{p}$. The second compartment of the frequency forms in the range of the lower values of $T$ and $\mu^{\star}$ and is related to the right hand part of some loops.

%
%
%

With the explicit form of the function $\Delta\left(\omega\right)$, one can calculate the exact value of the energy gap at the Fermi level. To do this, it is necessary to use the equation:

\begin{equation}
\label{r7}
\Delta\left(T\right)={\rm Re}\left[\Delta\left(\omega=\Delta\left(T\right)\right)\right].
\end{equation}

On the basis of the low-temperature solutions ($T=T_{0}$) and $\mu^{\star}\in\left<0.1,0.3\right>$ it has been stated, that $2\Delta\left(0\right)\in\left<50.96, 30.12\right>$ meV. Thus, the dimensionless ratio $R_{\Delta}\equiv 2\Delta\left(0\right)/k_{B}T_{C}$ assumes the high values from $4.55$ to $4.29$. Let us notice that the classical BCS theory predicts $\left[R_{\Delta}\right]_{{\rm BCS}}=3.53$ \cite{BCS}. From the physical point of view, the achieved result indicates that the thermodynamic properties of the superconducting state in 
SiH$_4$(H$_2$)$_2$ very significantly differ from the properties of the canonical superconducting state (the BCS theory).

In order to determine the values of $R_{\Delta}$ ratio it was necessary to conduct the complicated and time consuming calculations. For this reason, we present the analytical formula for $R_{\Delta}$, which allows to reconstruct the accurate numerical results: 

\begin{equation}
\label{r8}
\frac{R_{\Delta}}{\left[R_{\Delta}\right]_{{\rm BCS}}}=1+\left(\frac{k_{B}T_{C}}{a\omega_{{\rm ln}}}\right)^{2}
\left[\ln\left(\frac{a\omega_{{\rm \ln}}}{k_{B}T_{C}}\right)+\ln^{2}\left(\frac{a\omega_{{\rm \ln}}}{k_{B}T_{C}}\right)
\right],
\end{equation}
where $a=0.3447$.

The expression \eq{r8} has been constructed basing on $300$ exact values of the function $R_{\Delta}\left(\mu^{\star}\right)$ in the range of $\mu^{\star}$ from $0.1$ to $0.3$.

The second solution of the Eliashberg equations ($Z\left(\omega\right)$) serves as the base to calculate the exact value of the electron effective mass ($m^{\star}_{e}$). Performing the correspondent calculations, it has been found that the electron effective mass weakly depends on $T$ and $\mu^{\star}$. The quantity $m^{\star}_{e}$ takes its maximum value at the critical temperature: $m^{\star}_{e}=2.94 m_{e}$, where the symbol $m_{e}$ denotes the electron band mass. The achieved value has been calculated on the basis of the following expression: $m^{\star}_{e}/m_{e}={\rm Re}\left[Z\left(\omega=0,T=T_{C}\right)\right]$.

In the last step let us notice that the value of the electron effective mass accurately reproduces the value of the expression  $1.131\left(1+\lambda\right)$, that is included in the critical temperature formula (Eq. \eq{r6}).


In summary: we have precisely calculated the value of the critical temperature and the energy gap at the Fermi level for SiH$_4$(H$_2$)$_2$ compound under the influence of the pressure at $250$ GPa. We have examined the wide range of the Coulomb pseudopotential's values; $\mu^{\star}\in\left<0.1,0.3\right>$. It has been stated that $T_{C}$ can undergo the change from $129.83$ K to $81.40$ K. The obtained result means that even for the high value of $\mu^{\star}$ the critical temperature is very high. It should be underlined that the exact values of the critical temperature cannot be calculated with the help of the classical formulas.
   
In the considered range of the Coulomb pseudopotential's values, the low-temperature energy gap changes from $50.96$ meV to $30.12$ meV. The high values of $2\Delta\left(0\right)$ cause the dimensionless parameter $R_{\Delta}$ to greatly exceed the universal value $3.53$ predicted by the classical BCS theory. From the physical point of view, the discussed result indicates that the thermodynamic properties of the superconducting state in SiH$_4$(H$_2$)$_2$ significantly differ from the properties of the BCS superconducting state. 

In the presented paper, the analytical expressions for $T_{C}\left(\mu^{\star}\right)$ and $R_{\Delta}\left(\mu^{\star}\right)$ have been given. In the future, they will allow to obtain the physical Coulomb pseudopotential's value for SiH$_4$(H$_2$)$_2$ compound in the experimental way.

\begin{acknowledgments}
The authors wish to thank Prof. K. Dzili{\'n}ski for providing excellent working conditions and the financial support.
All numerical calculations were based on the Eliashberg function sent to us by: Prof. Yanming Ma and Prof. Yinwei Li for whom we are very thankful.\\
Some calculations have been conducted on the Cz{\c{e}}stochowa University of Technology cluster, built in the framework of the
PLATON project, no. POIG.02.03.00-00-028/08 - the service of the campus calculations U3.
\end{acknowledgments}
%

%
\end{document}